\documentclass[11pt]{article}
\begin{document}
\title
{A note on Hawking radiation via complex path analysis}
\author
{Sourav Bhattacharya{\footnote{sbhatt@bose.res.in }}\\
%\quad and Amitabha Lahiri{\footnote{amitabha@bose.res.in}}\\
S. N. Bose National Centre for Basic Sciences, \\
JD Block, Sector III, Salt Lake, Kolkata -700098, India.\\
}

\maketitle
%%%%%%%%%%%%%%%%%%%%%%%%%%%%%%%%%%%%%%%%%%%%%%%%%%%%%%%%%%%%%%%%%%%%%%%%%%%%%%%%%%%%%%%%%%%%%%%%%%%%%%%%%%%%%%%%%
\abstract

 {As long as we neglect backreaction, the Hawking temperature
of a given black hole would not depend upon the parameters of the
particle species 
we are considering. In the semiclassical
complex path analysis approach of Hawking radiation, this
has been verified by taking scalar and Dirac spinors 
separately for different stationary spacetime metrics. Here we show, in
a coordinate independent way that, for an arbitrary spacetime
with any number of dimensions, the equations of motion for a
Dirac spinor, a vector, spin-$2$ and spin-$\frac{3}{2}$ 
fields reduce to 
Klein-Gordon equations in the WKB semiclassical
limit. 
%The equation for a charged Dirac spinor reduce to that
%of a charged scalar.
% So, at the semiclassical level
%the scalar, spinor, vector and tensor equations of motion
%are equivalent in any arbitrary spacetime. 
%The equivalence of the equations of motion of different spin fields 
% shows that in the semiclassical
%approach we may compute the single particle emission probability
%and the Hawking temperature for different matter fields
 %through identical manner i.e., by considering the scalars only.
  We then obtain, under some suitable assumptions,
the complex solutions of those resulting scalar equations
across the Killing horizon of a stationary spacetime to get a
coordinate independent
expression for the emission probability identical  for
all particle species. Finally we consider some explicit
examples to demonstrate the validity of that expression. 
}

\hskip 1cm

    {\bf Keywords:}{ Hawking radiation,
 tunneling, spinor, vector 
%\maketitle

%\bigskip
\section{Introduction}
The semiclassical tunneling method 
\cite{Kraus:1994}-\cite{Paddy3:2002}
is an alternative approach to model particle creation by
black holes \cite{Hawk}. 
The basic scheme of this method is to compute the imaginary part of
the `particle' action which gives the emission probability from the 
event horizon. From the expression of the
 emission probability one identifies the 
temperature of the radiation. The earliest works in this context
can be found in \cite{Kraus:1994, Kraus:1996}. Following these works
an approach called the null geodesic method was developed 
\cite{Wilczek:2000, Parikh2:2004}. There exists also another way
to model black hole evaporation via tunneling called complex path
analysis \cite{Paddy1:1999, Paddy2:2001, Paddy3:2002} which we discuss here. This method involves writing down, in the 
semiclassical limit $\hbar \to 0$ a Hamilton-Jacobi equation 
from the matter equations of motion, treating the horizon as a 
singularity in the complex plane (which is a simple pole for all
known solutions) and then complex integrating the equation across
that singularity to obtain an imaginary contribution for the particle action. 

Both of this two alternative approaches have received great
attention during last few years. It is noteworthy that
since both of
these methods deal only with the near horizon geometry,
they can be
very useful alternatives particularly when the spacetime has no
well defined asymptotic structure or infinities \cite{sb}.

As far as we neglect the backreaction of the matter we are 
considering, the 
temperature of the radiation or the Hawking temperature should
not depend upon the parameters (e.g. mass, spin, and charge) of the 
particle species. The Smarr formula for black hole mechanics
predicts that this temperature is proportional to the surface gravity
of the event horizon for a stationary black hole with a Killing horizon. 

The complex path analysis approach has been successfully
applied to scalar emissions as well as to spinor emissions
separately for a wide class of stationary
black holes giving the expected expressions of Hawking temperatures
that were predicted by the Smarr formula. To tackle Dirac equation
in this approach the usual method has been employed, i.e., finding
a proper representation of the general $\gamma$ matrices in terms
of the Minkowskian $\gamma$'s and the metric functions and then
making the variable separation. 
For an exhaustive review and list of references on this
see e.g. \cite{rb}. See also e.g. \cite{Akhmedov:2006pg}-\cite{Belinski:2009bc} for some recent 
issues concerning the tunneling approach.

Thus, the universality of the Hawking temperature has been
proved case by case for a wide variety of black holes via the
 complex path method. Can we prove this universality from
a more general point of view?

In particular, in this paper we shall show that for the 
Dirac spinors we do not need 
to work with any
particular representation of the $\gamma$ matrices
in the semiclassical framework.  
In this work we wish to point out, in a coordinate 
independent way
that in any arbitrary spacetime with any number
of dimensions, the equations of motion for a
Dirac spinor, a vector, spin-$2$ meson
and spin-$\frac{3}{2}$ fields reduce to 
Klein-Gordon equations in the semiclassical
limit $\hbar \to 0$ for the usual WKB ansatz. 
The equations for a charged Dirac spinor reduce to that
of a charged scalar. This clearly shows that at the
semiclassical level all those different equations of motion
of various particle species are equivalent and it is
sufficient to deal with the scalar equation only. 
We shall also present, for a stationary spacetime with some
assumed geometrical properties, 
a general coordinate independent expression
for the emission probability and the Hawking temperature 
which is characterized by the black hole parameters itself (Eq. (\ref{e})).
We further consider some explicit examples to demonstrate
that our formula indeed gives the expected Hawking temperature
in terms of the horizon's surface gravity. 
 
Thus the semiclassical complex path method gives us a way
in which we may treat the different spin fields in an
identical footing, giving the same Hawking temperature
and thereby proving the universality of the Hawking 
temperature for stationary black holes from a very
 general point of view.

The paper is organized as follows. In the next section
 we shall deal with Dirac
spinors (neutral and then charged) to show that the equations
reduce to that of scalars in the semiclassical limit for the WKB
ansatz. In Sect. 3, we shall explicitly expand the 
resultant scalar equation in a coordinate independent
way in the near horizon limit for a stationary black hole
 with a Killing horizon, and shall
present a general expression that gives 
the emission or absorption probabilities. 
We shall illustrate the validity of this expression by
taking a few explicit examples.
In Sect. 4, we shall also
demonstrate that similar results hold also for
 the vector, massive spin-$2$ and spin-$\frac{3}{2}$ fields.
Finally we shall discuss our results. 

We shall take $G=1=c$, but shall retain $\hbar$ throughout.

%%%%%%%%%%%%%%%%%%%%%%%%%%%%%%%%%%%%%%%%%%%%%%%%%%%%%%%%%%%%%
\section{Reduction of the semiclassical Dirac equation into Klein-Gordon equation}

Let us then start by considering a spacetime of dimension $n$,
and a metric $g_{ab}$ defined on it, at least in our
region of interest.  We consider the Dirac equation
\begin{eqnarray}
i\gamma^a\nabla_a\Psi 
=\frac{m}{\hbar}\Psi.
\label{s1}
\end{eqnarray} 
$\nabla_a$ is the spin covariant derivative defined by $\nabla_a\Psi:=
\left(\partial_a+\Gamma_a\right)\Psi$, where $\Gamma_a$ are the spin connection matrices. The matrices
$\gamma^a(x)$ are the curved space generalization of the Minkowskian
$\gamma^{(\mu)}$. We expand 
$\gamma^a$ in an orthonormal basis, 
$\gamma^a=\gamma^{(\mu)}e_{(\mu)}^{a} :
\mu=0,~1,~2,\dots,~(n-1)$. Also, 
$g^{ab}e^{(\mu)}_{a}e^{(\nu)}_b=\eta^{(\mu)(\nu)}$. Here the Greek
indices within bracket denote the local Lorentz indices and 
$\eta^{(\mu)(\nu)}$ is the inverse metric corresponding to
the $n$-dimensional Minkowski spacetime. The 
$\gamma^{(\mu)}$ satisfy the well known anti-commutation relation:
$\left\{\gamma^{(\mu)},~\gamma^{(\nu)}\right\}=2 \eta^{(\mu)(\nu)}
\bf{I}$, where $\bf{I}$ denotes the identity matrix.
 
The expansion of $\gamma^a$ in terms of
the orthonormal basis $\{e_{(\mu)}^{a}\}$, and the
anti-commutation relation for $\gamma^{(\mu)}$'s give 
\begin{eqnarray}
\left\{\gamma^a,~\gamma^b\right\}=2g^{ab}\bf{I}.
\label{s2}
\end{eqnarray} 
Now we square Eq. (\ref{s1}) by acting with $i\gamma^b\nabla_b$
on both sides from left, producing
\begin{eqnarray}
\frac{1}{2}\left(\gamma^b\gamma^a+\gamma^a\gamma^b\right)
\nabla_b\nabla_a\Psi+
\frac{1}{4}\left(\gamma^b\gamma^a-\gamma^a\gamma^b\right)
\left(\nabla_b\nabla_a-\nabla_a\nabla_b\right)\Psi+
\left(\gamma^b\nabla_b \gamma^a\right)\nabla_a \Psi=
-\frac{m^2}{\hbar^2}\Psi.
\label{s3}
\end{eqnarray} 
But the commutator of two covariant derivatives acting
on $\Psi$ is proportional to the Riemann tensor,
$\left(\gamma^b\gamma^a-\gamma^a\gamma^b\right)
\left(\nabla_b\nabla_a-\nabla_a\nabla_b\right)\Psi=
\left(\gamma^a\gamma^b-\gamma^b\gamma^a\right)
R_{abcd}\left(\gamma^c\gamma^d-\gamma^d\gamma^c\right)
\Psi$.
Using this fact and 
 the anti-commutation relation for $\gamma^a$ 
(Eq. (\ref{s2})), Eq. (\ref{s3}) becomes 
\begin{eqnarray}
\nabla_a\nabla^a\Psi+
\frac{1}{4}\left[\gamma^a,~\gamma^b\right]
R_{abcd}\left[\gamma^c,~\gamma^d\right]\Psi+
\left(\gamma^b\nabla_b \gamma^a\right)\nabla_a \Psi=
-\frac{m^2}{\hbar^2}\Psi.
\label{s4}
\end{eqnarray} 
We will look at Eq. (\ref{s4}) semiclassically.  
We choose the usual WKB ansatz for a spin-`up' particle 
\begin{eqnarray}
\Psi  
&=&\left[ 
\begin{array}{c}
A(x) \\ 
0 \\ 
B(x) \\ 
0\\
\end{array}
\right] e^{\frac{i I(x)}{\hbar}}. 
\label{s5}
\end{eqnarray}
and substitute into Eq. (\ref{s4}).
Since we are neglecting backreaction, the 
components of the Riemann tensor are independent
of $\hbar$. Then
 it is clear that in the semiclassical
limit $\hbar \to 0$, 
 on the left 
hand side only the first term survives because
only this one contains some double derivatives of
${\cal {O}}\left(\hbar^{-2}\right)$. The
single derivative terms coming from the Laplacian
will certainly not survive in the semiclassical limit
(which is true for an actual
 scalar equation also), but we shall
formally keep the Laplacian $\nabla_a\nabla^a$ 
intact till later when we shall discuss its expansion
explicitly. Thus in the semiclassical limit, the WKB
ansatz (\ref{s5}) implies Eq. (\ref{s4})
can be effectively represented by two
Klein-Gordon equations for spin-`up' particles
\begin{eqnarray}
\nabla_a\nabla^a\Psi
+\frac{m^2}{\hbar^2}\Psi=0.
\label{s6}
\end{eqnarray} 
Similar result holds for a spin-`down' particle also.

If we consider
 a Dirac particle with a charge $e$
coupled to a gauge field $A_a$, the spin covariant 
derivative $\nabla_a$ in Eq. (\ref{s1}) is replaced
by the gauge covariant derivative 
$\widetilde{\nabla}_a \equiv \nabla_a-\frac{ie}{\hbar} A_a$ 
such that the equation of motion becomes
\begin{eqnarray}
i\gamma^a\nabla_a\Psi +\frac{e}{\hbar}\gamma^a A_a\Psi
=\frac{m}{\hbar}\Psi.
\label{s7}
\end{eqnarray} 
We now apply from the left
$\left(i\gamma^b\nabla_b+\frac{e}{\hbar}\gamma^b A_b\right)$
on both sides of this equation. Using Eq.s (\ref{s2}) and
(\ref{s4}) we obtain
\begin{eqnarray}
\nabla_a\nabla^a\Psi+
\frac{1}{4}\left[\gamma^a,~\gamma^b\right]
R_{abcd}\left[\gamma^c,~\gamma^d\right]\Psi+ 
\left(\gamma^b\nabla_b \gamma^a\right)\nabla_a \Psi 
-\frac{e^2}{\hbar^2}A_bA^b\Psi +\frac{2ie}{\hbar}A^a
\nabla_a\Psi\nonumber \\
-\frac{ie}{\hbar}\left[\left(\gamma^b\nabla_b \gamma^a\right)
A_a +\frac{1}{4}\left[\gamma^a,~\gamma^b\right]F_{ab}
 +\left(\nabla_a A^a\right)
\right]\Psi=-
\frac{m^2}{\hbar^2}\Psi,
\label{s8}
\end{eqnarray} 
where $F_{ab}=\nabla_a A_b-\nabla_b A_a$. We
now substitute
the ansatz (Eq. (\ref{s5})) into Eq. (\ref{s8})
and take the semiclassical limit $\hbar \to 0$.
 We see that in this 
limit Eq. (\ref{s8}) can formally be represented by
\begin{eqnarray}
\nabla_a\nabla^a\Psi 
-\frac{e^2}{\hbar^2}A_bA^b\Psi
+\frac{2ie}{\hbar}A^a\nabla_a\Psi+
\frac{m^2}{\hbar^2}\Psi=0,
\label{s9}
\end{eqnarray} 
each of 
which effectively has the form of the equation of motion of
a charged scalar.

What have we seen so far? We have dealt with
neutral and charged Dirac spinors and have
explicitly shown in a coordinate independent way
that, for the semiclassical WKB ansatz
all those equations of motion are equivalent
to that of scalars
in any arbitrary spacetime of dimension $n$. 
 So it is clear that the single particle
Hawking radiation will be
identical for Dirac spinors and scalars for any given
black hole.
% We shall give a general coordinate independent
%computation on the emission probability in the following part.

We shall also show explicitly in Sect. 4
that similar conclusions hold for Proca,  
massive spin-$2$ and spin-$\frac{3}{2}$ fields.
 But before that 
we wish to discuss the explicit expansions
and the near horizon limits of Eq.s 
 (\ref{s6}), (\ref{s9}) in a stationary
spacetime containing black hole. We shall address
only the charged Dirac spinor (or equivalently, charged scalar,
 Eq. (\ref{s9})). The other case will be equivalent to 
setting $e=0$ in Eq. (\ref{s9}).

%%%%%%%%%%%%%%%%%%%%%%%%%%%%%%%%%%%%%%%%%%%%%%%%%%%%%%%%%%%%%%%%%%%%%
\section{Hawking temperature for a stationary black hole with \\
Killing horizon}

We wish to present in the following a general 
coordinate independent expression
for the emission or absorption probability from a
stationary black hole with some assumed
geometrical properties. 
Let us first list the definitions and
 assumptions we make. 

We consider an $n$-dimensional stationary spacetime 
containing a black hole with a Killing
horizon ${\cal{H}}$.
We assume that the spacetime 
can be foliated into a family of hypersurfaces
$\Sigma$, orthogonal to a vector field $\chi^a$.
The hypersurface is spacelike everywhere except 
at the horizon (${\cal{H}}$), which is defined to be 
an $(n-1)$ dimensional null hypersurface. So, $\chi^a$ 
is orthogonal to a null hypersurface over ${\cal{H}}$
and hence $\chi^a$ is itself null over ${\cal{H}}$.
Everywhere else $\chi^a$ is timelike.

 Since ${\cal{H}}$ is a Killing horizon, the vector field
$\chi^a$ becomes a null Killing vector field, say
$\chi_{\rm{H}}^a$, over ${\cal{H}}$. $\chi^a$ is not
necessarily a Killing field everywhere, but it is 
Killing at least over ${\cal{H}}$ 
\begin{eqnarray}
\chi^a\vert_{{\cal{H}}}=\chi_{\rm{H}}^a:~\nabla_{(a}\chi_{{\rm{H}}b)}=0,~ \chi_{\rm{H}}^a\chi_{{\rm{H}}a}=-\beta^2\vert_{{\cal{H}}}=0.
\label{ad1}
\end{eqnarray} 
We now write the spacetime metric $g_{ab}$ as
\begin{eqnarray}
g_{ab}=-\beta^{-2}\chi_a\chi_b+\lambda^{-2}R_aR_b +\gamma_{ab},
\label{e1}
\end{eqnarray} 
where $R^a$ is a spacelike vector field orthogonal to $\chi^a$,
and $\lambda^2$ is the norm of $R_a$.
$\gamma_{ab}$ is the non-null spacelike portion of the metric 
perfectly well behaved on or in an infinitesimal neighbourhood 
of the horizon. 

Let us denote the Killing fields of this spacetime by
$(\xi_a,~\{\phi^{i}_a\})$, where $i=1,2\dots m$. Let $\xi_a$
be the timelike Killing field and $\{\phi^{i}_a\}$ be the
spacelike Killing field(s). We assume that the hypersurface
orthogonal vector field $\chi^a$ (which
is orthogonal to $\{\phi^{i}_a\}$ and 
any other spacelike field),
 can be written as a linear
combination of all the Killing fields
\begin{eqnarray}
\chi_a=\xi_a+\alpha^i(x)\phi^i_a,
\label{lm1}
\end{eqnarray} 
where repeated indices are summed over and $\{\alpha^i(x)\}$
are smooth functions. Then, using
Killing's equation we have $\nabla_{(a}\chi_{b)}
=\phi^i_{(a}\nabla_{b)}\alpha^i(x)$. Thus we have
\begin{eqnarray}
\chi^a\chi^b\nabla_a\chi_b=-\frac{1}{2}\chi^a\nabla_a\beta^2
=\chi^a\chi^b\phi^i_{a}\nabla_{b}\alpha^i(x)=0.
\label{lm2}
\end{eqnarray} 
Eq. (\ref{lm2}) shows that $\nabla_a\beta^2$ is
everywhere orthogonal to $\chi^a$ and hence it is
spacelike when $\chi^a$ is timelike. So, we may choose
$R_a=\nabla_a\beta^2$ in Eq. (\ref{e1}).

To look at the behaviour of $\nabla_a\beta^2$ over 
the horizon, we recall that 
over the Killing horizon ${\cal{H}}$
\cite{Wald:06,
 Gourgoulhon:2005ng}
\begin{eqnarray}
\nabla_a\beta^2=-2\kappa\chi_{{\rm{H}}a},
\label{g1}
\end{eqnarray} 
where $\kappa$ is a function. Since by definition 
$\chi_{\rm{H}}^a$ is null hypersurface orthogonal
at the horizon, it turns out that $\kappa$ is
a constant over the horizon \cite{Wald:06}. 
 Eq. (\ref{g1}) shows that $\nabla_a \beta^2$ is
 null over ${\cal{H}}$. 
%Thus we may choose $R^a$
%to be
%such that it
% smoothly coincides with $\nabla_a\beta^2$ as we
%reach ${\cal{H}}$
%
%\begin{eqnarray}
%R_a\stackrel {{\cal{H}}}\longrightarrow\nabla_a\beta^2.
%\label{ad2}
%\end{eqnarray} 
% 
 However, the choice $R_a=\nabla_a\beta^2$ is
not unique, we could have multiplied $
\nabla_a\beta^2$ by 
some non-diverging
function over ${\cal{H}}$, even some
positive power of $\beta$. But we shall retain
this choice for convenience.    

Let $R$ be the parameter along $R^a$. Then using Eq. 
(\ref{g1}) we have over ${{\cal{H}}}$ 
\begin{eqnarray}
R^a\nabla_a\beta^2=\frac{d\beta^2}{dR}
=-4\kappa^2\beta^2,
\label{g2}
\end{eqnarray} 
which implies over ${{\cal{H}}}$ 
\begin{eqnarray}
 \beta^2=e^{-4\kappa^2R}.
\label{g3}
\end{eqnarray} 
With the choice of $R^a$ we have made, it is clear
that the metric (\ref{e1}) becomes doubly
degenerate over ${{\cal{H}}}$. 
Note that Eq. (\ref{e1}) can readily be 
realized, in its doubly degenerate form, for
a static spherically symmetric black hole by employing
the usual $(t,~r_{\star})$ coordinates, where
$r_{\star}$ is the Tortoise coordinate. We shall be more
explicit about $R$ when we shall go into
specific examples.

 The assumption of stationarity and Killing horizon
would help us to provide 
a meaningful notion of the `particle' energy~\cite{Wald:06}. 

For $n>4$, the 
 uniqueness and other general properties of black
 holes 
are not very well understood and there may exist
 more general
 stationary black holes.
 However, we shall show below that for known
  stationary exact solutions,
 those assumptions will be sufficient.

Let us now expand Eq. (\ref{s9}) with the 
 decomposition (\ref{e1}). The single derivative terms do
not contribute in the $\hbar \to 0$ limit we are concerned
with
and the equation explicitly becomes
\begin{eqnarray}
\lambda^2\left(\chi^a\partial_a I-ef\right)^2
-\beta^2\left(R^a\partial_a I+
eg\right)^2-\left(\beta\lambda\right)^2\left[
\gamma_{ab}\partial^aI\partial^b I
+e^2\gamma_{ab}A^aA^b-
2e\gamma_{ab}A^a\partial^b I
+m^2\right]=0, %\nonumber\\
\label{g4}
\end{eqnarray} 
where $f=-\chi^aA_a$, and $g=R_aA^a$.
Here it is clear that had we multiplied $R^a$
by a function $h(x)$ non-diverging over 
${{\cal{H}}}$, we would have multiplied 
Eq. (\ref{g4}) only 
by an over all factor $h^2(x)$.

Now we shall look Eq. (\ref{g4}) in the near horizon
limit.
 By our
assumption the metric functions $\gamma_{ab}$ are well
behaved over the horizon. So, $\gamma_{ab}A^aA^b$ is
non divergent over ${{\cal{H}}}$.
Also, examples with
$g\neq 0$ seem to be unknown in the literature.
 So, we shall set $g=0$ in Eq. (\ref{g4}) and write Eq. (\ref{g4})
in the near horizon limit as
\begin{eqnarray}
\lambda^2\left(\chi^a\partial_a I-ef\right)^2
-\beta^2\left(R^a\partial_a I\right)^2-\left(\beta\lambda\right)^2\left[
\gamma_{ab}\partial^aI\partial^b I-
2e\gamma_{ab}A^a\partial^b I
\right]=0. 
\label{g5}
\end{eqnarray} 
To further simplify Eq. (\ref{g5}), let us choose an orthogonal
basis $\left\{m^a_{i}\right\}_{i=1}^{n-2}$ for $\gamma_{ab}$. Let
$\theta_i$ be the parameter along each $m^a_{i}$. Let us consider the
first term within the square brackets.
 This is basically a sum of the 
squares of $(n-2)$
 Lie derivatives: $\frac{1}{m_1^2}(\pounds_{m_1}I)^2+
\frac{1}{m_2^2}(\pounds_{m_2}I)^2\dots$, where $m_i^2$
is the norm of each $m^a_{i}$. By our definition, those
norms are non-zero finite over ${\cal{H}}$. Since $I$ is
a scalar those Lie derivatives are basically partial 
derivatives : $\pounds_{m_i}I=\partial_{\theta_i}I$. 

We shall now check whether the terms within the 
square bracket in Eq. (\ref{g5}) are divergent 
over ${\cal{H}}$. Let us 
suppose that close to ${{\cal{H}}}$, if possible the
following divergence occur
\begin{eqnarray}
\gamma_{ab}\partial^aI\partial^bI=\frac{D(x)}{\beta^2},
\label{g6}
\end{eqnarray} 
where $D(x)$ is bounded over or close to
${\cal{H}}$ and independent of
$\beta$ at leading order. 
 Then Eq. (\ref{g2}) implies
that $D(x)$ is also independent of $R$ over ${\cal{H}}$
\begin{eqnarray}
\pounds_{R}D(x)\Big\vert_{{\cal{H}}}=0.
\label{g7}
\end{eqnarray} 
Also by our choice $R_a=\nabla_a\beta^2$,
whose norm is $\lambda^2$, vanishes over ${\cal{H}}$ as
${\cal{O}}(\beta^2)$ (Eq. (\ref{g1})). So the function $D(x)$ is
also independent of $\lambda$ in the leading order
over ${\cal{H}}$.
Since the metric functions $\gamma_{ab}$ are well
behaved over ${\cal{H}}$, the
divergence of $\gamma_{ab}\partial^aI\partial^bI$ arises
from the Lie derivatives $(\partial_{\theta_i}I)^2$.
For simplicity we shall suppose that the divergence
comes from a single Lie derivative which is the $i$-th
one. We can easily generalize our analysis for more than
one diverging terms.
 Let us take near the horizon
\begin{eqnarray}
(\partial_{\theta_i}I)^2=\frac{C_{i}^2(x)}{\beta^2},
\label{g8}
\end{eqnarray} 
where $C_{i}^2(x)$ is a non-diverging function 
independent of $\beta$ in the leading order over 
or close to ${\cal{H}}$,
and is independent of $R$ over
 ${\cal{H}}$. 

The divergence of the second term within the
 square bracket in Eq. (\ref{g5})
comes from $(\partial_{\theta_i}I)$ which, by 
Eq. (\ref{g8}) is ${\cal{O}}(\beta^{-1})$. So this
term can be neglected with respect to the quadratic
term $(\partial_{\theta_i}I)^2$. Hence
comparing Eq.s (\ref{g6}),
 (\ref{g8}) we have $D(x)=\frac{C_{i}^2(x)}{m_i^2}$.

 Using Eq. (\ref{g2}) 
we obtain from Eq. (\ref{g8}) the following divergence over 
${\cal{H}}$
\begin{eqnarray}
\frac{\partial^2I}{\partial R\partial{\theta_i}}=\pm\frac{2\kappa^2
C_{i}(x)}{\beta}.
\label{g9}
\end{eqnarray} 
On the other hand 
we can write Eq. (\ref{g5}) near ${\cal{H}}$ now as
\begin{eqnarray}
\left(\partial_R I\right)^2=\frac{\lambda^2}{\beta^2}\left[\left(\chi^a\partial_a I-ef\right)^2-D(x)\right].
\label{e6}
\end{eqnarray} 
We shall take the Lie derivative of Eq. (\ref{e6}) with
respect to $m_i^a$ over ${\cal{H}}$. 
By our choice $R_aR^a=\lambda^2=\nabla_a \beta^2
\nabla^a\beta^2$. Also, the function $\kappa$
in Eq. (\ref{g1}) is a constant over ${\cal{H}}$. This means 
that $\partial_{\theta_i}\kappa=0$ over ${\cal{H}}$.
Since by our definition the vector
field  
$\chi_{\rm{H}}^a$ is Killing  
over ${\cal{H}}$, the term
 $\left(\chi_{\rm{H}}^a\partial_a I-ef\right)$ is a conserved
quantity, 
 i.e., a constant \cite{Wald:06}.
We shall regard this term to be the conserved
effective energy $(E)$ of the particle. 
 So, using Eq.s
(\ref{g1}), (\ref{g9})
 the Lie derivative of Eq. (\ref{e6}) 
with respect to $m_i^a$ gives the
following ${\cal{O}}(\beta^{-1})$ 
divergence over ${\cal{H}}$
\begin{eqnarray}
\partial_{\theta_i}D(x)
=\pm \frac{\lambda}{\beta^2} C_i(x)\left[
E^2-
D(x)\right]^{\frac{1}{2}}.
\label{e8}
\end{eqnarray} 
Eq. (\ref{e8}) contradicts the fact that $D(x)$ is independent
of $\beta$, $\lambda$ or $R$ in the leading order over ${\cal{H}}$.
So, Eq. (\ref{g6}) cannot be true.    
Similarly we can show that
the term $\gamma_{ab}\partial^aI\partial^bI$
 cannot be divergent as ${\cal{O}}(\beta^{-n})$ 
for any $n>2$.
 Thus $\beta^2\gamma_{ab}\partial^aI\partial^bI=0$
over the horizon.

With all these, we now 
integrate Eq. (\ref{g5}) across the horizon along a complex path
\begin{eqnarray}
I_{\pm} 
=\pm\int_{{\cal{H}}}
\frac{\lambda}{\beta}\left(\chi_{\rm{H}}^a\partial_a I-ef\right)
dR,
\label{e}
\end{eqnarray} 
where complex integration is understood. The $+(-)$ sign
stands for outgoing (incoming) solution.
 Eq. (\ref{e})
gives the emission (absorption)
 probability for a
stationary black hole satisfying the assumptions we have made.

In order to verify the validity of
 Eq. (\ref{e}), at this point 
we need some particular metrics. 
We shall find out the vector fields 
 $\chi_{\rm{H}}^a$ and $R^a$, and then
compute $I_{\pm}$ from
Eq. (\ref{e}).

 Let us start with four dimensions by considering 
the charged Kerr black hole
\begin{eqnarray}
ds^2=-\frac{\Delta-a^2\sin^2\theta}{\Sigma}dt^2
-\frac{2a\sin^2\theta\left(r^2+a^2-\Delta\right)}{\Sigma}
dtd\phi&+& \frac{\left(r^2+a^2\right)^2-
\Delta a^2 \sin^2\theta}{\Sigma}\sin^2\theta d\phi^2 \nonumber \\
&+&\frac{\Sigma}{\Delta}dr^2+\Sigma d\theta^2,
\label{e10}
\end{eqnarray} 
where $\Sigma=r^2+a^2\cos^2\theta$, $\Delta=r^2+a^2+Q^2-2Mr$
; $a$ and $Q$ are the parameters specifying
rotation and charge respectively. $\Delta=0$ defines the horizon
($r_{\rm{H}}$). The gauge field of this solution is
$A_a=-\frac{Qr}{\Sigma}\left[(dt)_a-a\sin^2\theta
(d\phi)_a\right]$.

We first define 
$\chi^a=(\partial_{t})^a-\frac{g_{t\phi}}{g_{\phi\phi}}
(\partial_{\phi})^a$, such that $\chi_a(\partial_{\phi})^a=0$
everywhere. 
Near the horizon we have $\chi_a\chi^a=-\beta^2\approx-\frac{\Delta \Sigma}{\left(r^2+a^2
\right)^2-\Delta a^2 \sin^2\theta}\leq 0$. So, $\beta^2=0$ over the
horizon which implies $\chi^a$ becomes null over the horizon and timelike
outside it. 

Over the horizon $\chi^a$ becomes,
$\chi_{\rm{H}}^a=(\partial_{t})^a-\frac{g_{t\phi}}{g_{\phi\phi}}
(r_{\rm{H}})(\partial_{\phi})^a=(\partial_{t})^a+
\frac{a}{r_{\rm{H}}^2+a^2}(\partial_{\phi})^a
$, which is Killing and null. Thus we have specified
the required vector field $\chi^a$ which becomes null
and Killing over the horizon. 

 Next we need to find out $R^a$ and the parameter
 $R$ along it. Using the expression $\chi^a=(\partial_{t})^a-\frac{g_{t\phi}}{g_{\phi\phi}}(\partial_{\phi})^a$, we have $\chi^a\nabla_a\beta^2=0$
everywhere. So we can
let $R_a=\nabla_a\beta^2$. Then
using the expressions for $\beta^2$ and the metric functions
 (Eq. (\ref{e10})) we have near the horizon
\begin{eqnarray}
R^a\nabla_a\beta^2=\frac{d\beta^2}{dR}=\nabla_a\beta^2\nabla^a\beta^2
=\frac{\Delta \Sigma}{(r^2+a^2)^4}\Delta^{\prime2}+ {\cal{O}}(\Delta^2),
\label{e9}
\end{eqnarray} 
where the prime denotes derivative with respect to $r$.
Thus we have
found out the norm $\lambda^2(=\nabla_a\beta^2\nabla^a\beta^2)$ of
the vector field $R^a$ which becomes null over the horizon.
Also. Eq. (\ref{e9}) gives near the horizon
\begin{eqnarray}
R=\int\frac{(r^2+a^2)^2d\Delta}{\Delta\Delta^{\prime2}}.
\label{null}
\end{eqnarray} 
Thus we have specified the coordinate or the parameter $R$
along $R_a$. Note that Eq. (\ref{null}) implies that near
the horizon, choosing the vector field $R_a=\nabla_a\beta^2$
means a coordinate transformation $r\to R$ in the 
metric (\ref{e10}).

The components of the gauge field $A_a$ on the horizon are given 
by $A_{a}\chi_{\rm{H}}^a=-\frac{Q r_{\rm{H}}}{r_{\rm{H}}^2+a^2}$,
and $A_a(\partial_{\phi})^a=\frac{Qr_{\rm{H}}a^2\sin^2\theta}
{\left(r_{\rm{H}}^2+a^2\cos^2\theta\right)\left(r_{\rm{H}}^2+
a^2\right)}$. The near horizon contribution comes only from
the first one.

Substituting the near horizon norms
$\chi_a\chi^a=-\beta^2\approx-\frac{\Delta \Sigma}{\left(r^2+a^2\right)^2}$, $R^aR_a=\lambda^2=\frac{\Delta \Sigma}{(r^2+a^2)^4}\Delta^{\prime2}$, and
$dR=\frac{(r^2+a^2)^2d\Delta}{\Delta\Delta^{\prime2}}$ 
into Eq. (\ref{e}) we have 
\begin{eqnarray}
I_{\pm} 
=\pm\int_{{\cal{H}}}
\left(\chi_{\rm{H}}^a\partial_a I-ef\right)\frac{r^2+a^2}{\Delta}
dr,
\label{e11}
\end{eqnarray} 
where $f=-A_{a}\chi_{\rm{H}}^a=-
\frac{Q r_{\rm{H}}}{r_{\rm{H}}^2+a^2}$.
Eq. (\ref{e11}) was first obtained in
 \cite{Kerner:2008qv, Li:2008zra} by explicitly
solving the semiclassical
Dirac equation by method of separation of variables. 

The emission (absorption) probabilities are
given by $\sim \Big \vert  e^{\frac{iI(r_{\rm{H}})_{\pm}}{\hbar}}
\Big\vert^2$ \cite{Paddy1:1999}. We shall not go into the details
of the complexification of the `path', 
the choice of contours and explicit evaluation of
 Eq. (\ref{e11}). We refer the reader to
 \cite{Paddy1:1999, Kerner:2008qv,
Li:2008zra} for this.
 Explicit evaluation of Eq. (\ref{e11}) and the emission
($P_{\rm{E}}$) or
absorption ($P_{\rm{A}}$)
probabilities give the desired
temperature of the emission
from the exponential
behaviour of $\frac{P_{\rm{E}}}{P_{\rm{A}}}$.
 The Hawking temperature is found to be
 $T_{\rm{H}}=\frac{\kappa_{\rm{H}}}{2\pi}$, where
$\kappa_{\rm{H}}$ is the surface gravity of the event horizon.

After this, we shall consider some examples from higher dimensions. 
First, we consider non-extremal rotating charged black hole solution
of five dimensional minimal supergravity with two different
rotation parameters in the Boyer-Lindquist 
coordinates \cite{Chong:2005hr},
\begin{eqnarray}
ds^2 = &-&\left[\frac{\Delta_{\theta}\left(1+g^2r^2\right)}{\Sigma_a\Sigma_b}-
\frac{\Delta_{\theta}^2\left(2m\rho^2-q^2+2abqg^2\rho^2
\right)}{\rho^4\Sigma_a^2\Sigma_b^2}
\right]dt^2+\frac{\rho^2}{\Delta_r}dr^2+\frac{\rho^2}{\Delta_\theta}
d\theta^2 \nonumber\\
&+& \left[\frac{\left(r^2+a^2\right)\sin^2\theta}{\Sigma_a}+
\frac{a^2 \left(2m\rho^2-q^2\right)\sin^4\theta +2abq\rho^2\sin^4\theta }{\rho^4 \Sigma_a^2}\right] d\phi^2 \nonumber\\
&+&\left[\frac{\left(r^2+b^2\right)\cos^2\theta}{\Sigma_b}+
\frac{b^2 \left(2m\rho^2-q^2\right)\cos^4\theta+2abq\rho^2\cos^4\theta }
{\rho^4 \Sigma_b^2}\right] d\psi^2\nonumber\\
&-&\frac{2\Delta_{\theta}\sin^2\theta\left[a\left(2m\rho^2-q^2\right)
+ bq\rho^2\left(1+a^2g^2\right) \right]}
{\rho^4\Sigma_a^2\Sigma_b}dtd\phi \nonumber\\
&-&\frac{2\Delta_{\theta}\cos^2\theta\left[b\left(2m\rho^2-q^2\right)
+ aq\rho^2\left(1+b^2g^2\right) \right]}
{\rho^4\Sigma_a\Sigma_b^2}
dtd\psi\nonumber\\
&+&\frac{2\sin^2\theta\cos^2\theta\left[ab\left(2m\rho^2-q^2\right)
+ q\rho^2\left(a^2+b^2\right) \right]}
{\rho^4\Sigma_a\Sigma_b}
d\phi d\psi,%\nonumber\\
\label{e12}
\end{eqnarray} 
where $\rho^2=\left(r^2+a^2\cos^2\theta+b^2\sin^2\theta\right)$, 
$\Delta_{\theta}=\left(1-a^2g^2\cos^2\theta-b^2g^2\sin^2\theta
\right)$, $\Sigma_a=(1-a^2g^2)$, $\Sigma_b=(1-b^2g^2)$ and
$\Delta_r=\left[\frac{(r^2+a^2)(r^2+b^2)
(1+g^2r^2)+q^2+2abq}{r^2}-2M\right]$.
The black hole event horizon $(r_{\rm{H}})$ is
 given by $\Delta_r(r_{\rm{H}})=0$. The parameters
$(M,~a,~b,~q)$ specify respectively 
the mass, angular momenta and the charge of the black hole. $g$ is
a real positive constant. The gauge field corresponding 
to the charge $q$ is given by $A_a=
\frac{\sqrt{3}q}{\rho^2}\left(\frac{\Delta_{\theta}}{\Sigma_a
\Sigma_b}(dt)_a-\frac{a\sin^2\theta}{\Sigma_a}(d\phi)_a 
-\frac{b\cos^2\theta}{\Sigma_b}(d\psi)_a\right)$. 

The angular velocities of the comoving observers on the horizon
are given by \cite{Li:2010zzd}
\begin{eqnarray}
\Omega_{\phi}=- \frac{\left\{g_{t\phi}g_{\psi\psi}-g_{t\psi}g_{\phi\psi}\right\}}
{\left\{g_{\phi\phi}g_{\psi\psi}-(g_{\psi\phi})^2\right\}}
\Bigg\vert_{r=r_{\rm{H}}}
=\frac{a(r_{\rm{H}}^2+b^2)(1+g^2r_{\rm{H}}^2)+bq}
{(r_{\rm{H}}^2+a^2)(r_{\rm{H}}^2+b^2)+abq}, \nonumber\\
\Omega_{\psi}=-\frac{\left\{g_{t\psi}g_{\phi\phi}-g_{t\phi}g_{\phi\psi}\right\}}
{\left\{g_{\phi\phi}g_{\psi\psi}-(g_{\psi\phi})^2\right\}}
\Bigg\vert_{r=r_{\rm{H}}}=
\frac{b(r_{\rm{H}}^2+a^2)(1+g^2r_{\rm{H}}^2)+aq}
{(r_{\rm{H}}^2+a^2)(r_{\rm{H}}^2+b^2)+abq}.
\label{e13}
\end{eqnarray} 
We note that the vector field
\begin{eqnarray}
\chi^a 
=(\partial_{t})^a-\frac{\left\{g_{t\phi}g_{\psi\psi}-g_{t\psi}g_{\phi\psi}\right\}}
{\left\{g_{\phi\phi}g_{\psi\psi}-(g_{\psi\phi})^2\right\}}
(\partial_{\phi})^a
-\frac{\left\{g_{t\psi}g_{\phi\phi}-g_{t\phi}g_{\phi\psi}\right\}}
{\left\{g_{\phi\phi}g_{\psi\psi}-(g_{\psi\phi})^2\right\}}(\partial_{\psi})^a
\label{5d1}
\end{eqnarray} 
is orthogonal to $(\partial_{\phi})^a$ and $(\partial_{\psi})^a$ everywhere.
Also, the near horizon norm of $\chi^a$ is
$\chi^a\chi_a
=-\beta^2=-\frac{\rho^2r^4\Delta_r} {\left[(r^2+a^2)(r^2+b^2)+abq
\right]^2}+{\cal{O}}({\Delta_r^2})$. Thus $\chi^a$ becomes null over
the horizon.

Also, 
Eq. (\ref{e13}) shows that
 $\chi^a$ becomes a Killing field $\chi_{\rm{H}}^a$
 over the horizon, where
\begin{eqnarray}
\chi_{\rm{H}}^a 
=(\partial_{t})^a+\Omega_{\phi}(\partial_{\phi})^a
+\Omega_{\psi}(\partial_{\psi})^a.
\label{e14}
\end{eqnarray} 
  So, we have specified the
required vector field $\chi^a$ which becomes
null and Killing over the horizon. 

 Also, exactly
through the same manner as in the Kerr-Newman metric, we can specify
the other null vector field $R^a$, its norm $\lambda^2$,
and the coordinate $R$ for the metric (\ref{e12}). Choosing
$R_a=\nabla_a\beta^2$ we have near the horizon
\begin{eqnarray}
R_aR^a=\lambda^2=\nabla_a\beta^2\nabla^a\beta^2=R^a\nabla_a\beta^2
=\frac{\Delta_r\rho^2r^4}
{\left[(r^2+a^2)(r^2+b^2)+abq
\right]^4}\left(r^2\Delta_r\right)^{\prime 2},
\label{5d2}
\end{eqnarray} 
which becomes null over the horizon. Also, near the horizon
the coordinate $R$ along $R^a$ is given by
\begin{eqnarray}
R=\int \frac{\left[(r^2+a^2)(r^2+b^2)+abq
\right]^2 d(r^2\Delta_r)}{(r^2\Delta_r)(r^2\Delta_r)^{\prime 2}},
\label{5d3}
\end{eqnarray} 
The gauge field $A_a$ has three components :
$(A_a\chi_{\rm{H}}^a,~A_a(\partial_{\phi})^a,~
A_a(\partial_{\psi})^a)$, of which the near horizon
contribution comes only from $A_a\chi_{\rm{H}}^a$.

Substituting the near horizon norms $\beta^2=
\frac{\rho^2r^4\Delta_r} {\left[(r^2+a^2)(r^2+b^2)+abq
\right]^2}$, $\lambda^2=\frac{\Delta_r\rho^2r^4}
{\left[(r^2+a^2)(r^2+b^2)+abq
\right]^4}\left(r^2\Delta_r\right)^{\prime 2}$ 
and $dR=\frac{\left[(r^2+a^2)(r^2+b^2)+abq
\right]^2 d(r^2\Delta_r)}{(r^2\Delta_r)(r^2\Delta_r)^{\prime 2}}$ 
into Eq. (\ref{e}) we have  
\begin{eqnarray}
I_{\pm} 
=\pm\int_{{\cal{H}}}
\left(\chi_{\rm{H}}^a\partial_a I-ef\right)\frac{\left[
(r^2+a^2)(r^2+b^2)+abq\right]}
{r^2\Delta_r}dr,
\label{e15}
\end{eqnarray} 
where $f=-A_a\chi_{\rm{H}}^a=-\frac{\sqrt{3}q r_{\rm{H}}}
{(r^2+a^2)(r^2+b^2)+abq}$.
 Eq. (\ref{e15}) was first obtained in \cite{Li:2010zzd}
by explicit solution of
the semiclassical Dirac equation by method of separation
of variables.
Complex integration of Eq. (\ref{e15}) across the horizon
 and computation of the emission (absorption) probabilities
give the expected Hawking temperature in terms of the Killing
horizon's surface gravity \cite{Chong:2005hr, Li:2010zzd}.

 It can be easily verified using the same methods
as above
 that Eq. (\ref{e}) also applies
well and recovers the desired results for
 the $5$ dimensional stationary solutions with
Killing horizons like
 Kerr-G\"{o}del black hole \cite{Hashimoto:2003},
 squashed Kaluza-Klein black hole 
\cite{Kurita:2008mj, Ishihara:2005dp}, 
%squashed Kerr-G\"{o}del black hole \cite{Tomizawa:2008hw},
a black string \cite{Kurita:2008mj, Horowitz:2002ym},
black hole solutions of $z = 4$ Horava-Lifshitz gravity
\cite{Park:2009zra, Chen:2009bja} and the toroidal black
hole solutions like in \cite{Rinaldi:2002tc}.

Our scheme also applies very easily to an $n$
dimensional generalization of the Kerr black hole
with a single rotation parameter
\cite{Myers:1986un} 
\begin{eqnarray}
ds^2 &=&-dt^2+(r^2+a^2)\sin^2\theta d\phi^2+\frac{\mu}{
r^{n-5}\Sigma}\left(dt-a\sin^2\theta d\phi\right)^2+
\frac{r^{n-5} \Sigma} {r^{n-5}(r^2+a^2)-\mu}dr^2\nonumber\\
&+& \Sigma d\theta^2+r^2\cos^2\theta d\Omega^{n-4},
\label{e16}
\end{eqnarray} 
where the parameters $(\mu,~a)$ represents the mass and
angular momentum of the black hole. 
$\Sigma=r^2+a^2\cos^2\theta$ and $d\Omega^{n-4}$ represents
the metric over an $(n-4)$ sphere.

Eq. (\ref{e}) applies to a de Sitter horizon also, provided
the assumptions stated at the beginning of this section are
true for that case. Such an example is the Kerr-de Sitter
spacetime. The de Sitter horizon for this spacetime is a
Killing horizon \cite{Gibbons:1977mu}. One can show,
following exactly the similar way as before that all the 
other assumptions are valid for this case. Explicit
evaluation of Eq. (\ref{e}) gives the expected thermal
character of the incoming radiation.

%\begin{eqnarray}
%ds^2 & = & -u(r)dt^2-2g(r)\left( \cos \theta~ d \phi dt+  d \psi dt \right)+\frac{r^4}{\Delta(r)} dr^2+
% 2 \left(h(r)+\frac{r^2}{4}\right)\cos \theta d \phi d \psi \nonumber\\ 
%&& + \frac{r^2}{4}~d\theta^2 + \left(h(r) \cos^2 \theta + \frac{r^2}{4}\right)d \phi^2+\left(h(r) + \frac{r^2}{4}\right)d \psi^2,
%\label{e16} 
%\end{eqnarray}
%where
%$u(r)=1-\frac{2 M}{r^2}$, $g(r)=jr^2+\frac{M a}{r^2}$, 
%$h(r)=-j^2 r^2(r^2+2 M)+\frac{M a^2}{2 r^2}$,
%$\Delta(r)= r^4-2 Mr^2+ 8 j M(a+2j M)r^2+2 M a^2$.
%The parameters $(M,~j,~a)$ represents respectively
%the mass, rotation of the background and the rotation
%of the black hole \cite{Hashimoto:2003}. The gauge
%field is given by 
%$A_a=\frac{\sqrt{3}}{2}jr^2 \left[(d \psi)_a+\cos{\theta}
%(d \phi)_a\right]$. The black hole event horizon ($r_{\rm{H}}$)
% is given by $\Delta(r_{\rm{H}})=0$. The solution (\ref{e16})
%possesses closed timelike curves for $g_{\psi\psi}\leq0$
%\cite{Hashimoto:2003}. We shall work in the small $j$
%approximation (i.e. ${\cal{O}}(j^2)$ terms will be 
%neglected for finite $r$) in order push those curves off to
%large distances \cite{Kerner:2007jk}. 
% The Killing field $\chi_{\rm{H}}^a:=
%((\partial_{t})^a-\frac{g_{t\psi}}{g_{\psi\psi}}
%\Big\vert_{r_{\rm{H}}}(\partial_{\psi})^a)$ 
%has a norm $-\beta^2=-\frac{\Delta}{r^4+2Ma^2}$.

%%%%%%%%%%%%%%%%%%%%%%%%%%%%%%%%%%%%%%%%%%%%%%%%%%%%%%%%%%%%%%%%%%%%%%%%%%%%%%%%%%%%%%%%%%%%
\section{Vector, spin-$2$ and spin-$\frac{3}{2}$ fields}

Now we shall show that all the approaches and conclusions made in the
preceding sections
also hold for the Proca, massive spin-$2$ and spin-$\frac{3}{2}$ 
fields. Let us first consider the equation of motion for a Proca field $A^b$,
\begin{eqnarray}
\nabla_{a}F^{ab} = -\frac{m^{2}}{\hbar^2} A^{b},
\label{v1}
\end{eqnarray} 
where $F_{ab}=\nabla_a A_b-\nabla_b A_a$. Eq. (\ref{v1}) can be
written as 
\begin{eqnarray}
\nabla_a\nabla^a A_b -R_{b}{}^{a}A_a-
\nabla_b\left(\nabla_a A^a\right)=-\frac{m^{2}}{\hbar^2} A_b.
\label{v2}
\end{eqnarray} 
But Eq. (\ref{v1}) implies that $\nabla_a A^a=0$ identically.
%For $m=0$ in Eq. (\ref{v1}) we may take $\nabla_a A^a=0$ as a gauge.
Now let us choose a set of orthonormal basis $\left\{e_{a}^{(\mu)}\right\}$. We expand the vector field $A_a$ 
in this basis, $A_b=e_{b}^{(\mu)}A_{(\mu)}$. With this
expansion and the fact that $\nabla_a A^a=0$, Eq. (\ref{v1})
becomes
\begin{eqnarray}
e_{b}^{(\mu)}\nabla_a\nabla^a A_{(\mu)}+ A_{(\mu)}\nabla_a\nabla^a 
e_{b}^{(\mu)}+2\nabla_a A_{(\mu)}\nabla^a e_{b}^{(\mu)} 
-R_{b}{}^{(\mu)}A_{(\mu)}
=-\frac{m^{2}}{\hbar^2}A_{(\mu)}e_{b}^{(\mu)},
\label{v3}
\end{eqnarray} 
which, after contracting both sides by $e^{b}_{(\nu)}$, reduces
to   
\begin{eqnarray}
\nabla_a\nabla^a A_{(\nu)}+ A_{(\mu)}e^{b}_{(\nu)}
\nabla_a\nabla^a 
e_{b}^{(\mu)}+2e^{b}_{(\nu)}\nabla_a A_{(\mu)}\nabla^a 
e_{b}^{(\mu)} 
-R_{(\nu)}{}^{(\mu)}A_{(\mu)}
=-\frac{m^{2}}{\hbar^2} A_{(\nu)}.
\label{v4}
\end{eqnarray} 
 We
choose the usual WKB ansatz for each $A_{(\nu)}$ :
$A_{(\nu)}=f_{\nu}(x) e^{\frac{i I(x)}{\hbar}}$, substitute
into Eq. (\ref{v4}), and take the semiclassical limit $\hbar
\to 0$. 

Then it immediately turns out that
in the semiclassical limit Eq. (\ref{v4}) can be
effectively represented by $n$ 
 Klein-Gordon equations for the scalars $A_{(\nu)}$
\begin{eqnarray}
\nabla_a\nabla^a A_{(\nu)}
+\frac{m^{2}}{\hbar^2} A_{(\nu)}=0,
\label{v5}
\end{eqnarray} 
with $\nu=0,~1,~2,\dots,~(n-1)$. When each of the 
Eq.s (\ref{v5}) is
explicitly expanded and the near horizon limit is taken,
we get back Eq. (\ref{e}) with $e=0$.

Next, we turn our attention to the massive spin-$2$ field $\pi_{ab}$
satisfying Pauli-Fierz equation \cite{pauli}
\begin{eqnarray}
\nabla_c\nabla^c \pi_{ab}+\frac{m^2}{\hbar^2}\pi_{ab}=0,
\label{h1}
\end{eqnarray} 
where $\pi_{ab}$ are symmetric tensor fields. As before we 
expand $\pi_{ab}$ in orthonormal basis, 
$\pi_{ab}=e^{(\mu)}_a e^{(\nu)}_b\pi_{(\mu)(\nu)}$. In the
semiclassical limit and for the WKB ansatz
 Eq. (\ref{h1}) can effectively be represented
by $\frac{n(n+1)}{2}$ Klein-Gordon equations for the 
scalars $\pi_{(\mu)(\nu)}$ 
\begin{eqnarray}
\nabla_c\nabla^c \pi_{(\mu)(\nu)}+
\frac{m^2}{\hbar^2}\pi_{(\mu)(\nu)}=0,
\label{h2}
\end{eqnarray} 
and thus similar conclusions hold for this case also.

             %%%          SPIN 3/2

Finally, we wish to briefly address the spin-$\frac{3}{2}$ fields satisfying the Rarita-Schwinger equation \cite{rarita}. The tunneling phenomenon for this field was addressed in \cite{Yale:2008kx}
for the Kerr black hole by explicitly solving the equations of motion in the near horizon limit.

 The Rarita-Schwinger equation in a curved spacetime reads
\begin{eqnarray}
i\gamma^a\nabla_a\Psi_b =\frac{m}{\hbar}\Psi_b,
\label{h3}
\end{eqnarray} 
where $\Psi_b \equiv \Psi_b^{(s)}$ is a spinor with $s$ being
the spin index. The $\gamma$'s are matrices (with matrix indices
suppressed) satisfying the anti-commutation relation similar to the
Dirac $\gamma$'s: $\left\{\gamma^a,~\gamma^b \right\}=2g^{ab}\bf{I}$.
 The spin-covariant derivative $\nabla$ is defined as $\nabla_a \Psi_b:=\left(\partial_a + \Gamma_a\right)\Psi_b$, where 
$\Gamma_a$ are the spin connection matrices (with suppressed matrix indices). Also, $\Psi_b$ satisfies an additional
constraint $\gamma^a\Psi_a=0$.

Due to the similarity of the spin-$\frac{3}{2}$ fields with the Dirac spinors discussed in Sect. 2, we shall apply the same method here to show
that $\Psi_b$ satisfies the Klein-Gordon equation in the semiclassical WKB
framework.
 So, we square Eq. (\ref{h3}) by applying $i\gamma^c\nabla_c$ from
left. A little computation, using the definition of the spin-covariant derivative $\nabla_a$, the anti-commutation relation satisfied by the $\gamma$'s, and also the commutativity of the partial derivatives yields
\begin{eqnarray}
\nabla_a\nabla^a\Psi_b+
\frac{1}{4}\left[\gamma^a,~\gamma^c\right]
 \left(\partial_{[a}\Gamma_{c]}+\Gamma_{[a}\Gamma_{c]}\right)\Psi_b+
\left(\gamma^c\nabla_c \gamma^a\right)\nabla_a \Psi_b=
-\frac{m^2}{\hbar^2}\Psi_b.
\label{h4}
\end{eqnarray} 
So, as in the previous cases, it immediately follows then for the usual ansatz
\begin{eqnarray}
\Psi_a  
&=&\left[ 
\begin{array}{c}
A_a(x)e^{\frac{i I_1(x)}{\hbar}}\\ 
B_a(x)e^{\frac{i I_2(x)}{\hbar}} \\
C_a(x)e^{\frac{i I_3(x)}{\hbar}} \\
D_a(x)e^{\frac{i I_4(x)}{\hbar}}\\
\end{array}
\right], 
\label{h5}
\end{eqnarray}
Eq. (\ref{h4}) reduce to the Klein-Gordon equations in the semiclassical limit. We can easily generalize this result for a charged spin-$\frac{3}{2}$ particle coupled to a gauge field by replacing the spin covariant derivative by the gauge spin covariant derivative. This gives 
charged Klein-Gordon equations.

%%%%%%%%%%%%%%%%%%%%%%%%%%%%%%%%%%%%%%%%%%%%%%%%%%%%%%%%%%%%%%%%%%%%%%%
\section{Discussions}

We now summarize our results. The objective
of this work was to put the complex path approach
for stationary black holes
in a general framework. To do this,
 we have dealt with some well known physical
matter equations and shown for any arbitrary spacetime
 in a coordinate independent way that in the semiclassical
limit the WKB ansatz implies that all those
equations of motion are equivalent to the Klein-Gordon
equation.
We have done this without choosing any particular 
basis of the vector fields or the $\gamma$ matrices. 
We needed to assume only that a metric $g_{ab}$ can be defined
on the spacetime which guarantees 
the existence of the orthonormal basis $\left\{e^{(\mu)}_a
\right\}$ \cite{Wald:06}. So it is clear 
that as far as the semiclassical
level is concerned it is sufficient to work only with scalars
for any arbitrary black hole. It also becomes clear from
that the Hawking temperature is indeed independent of
the particle species we are concerned with.

 We further presented a general coordinate independent
expression for the emission probability from an
arbitrary stationary black hole with some assumed geometrical
properties
(Eq. ({\ref{e})). We showed that
finding the emission probability or the Hawking temperature
for such black holes reduces to merely finding  
a null coordinate or a null vector field
(which is spacelike outside
the horizon), and the norm of
the timelike vector field which is orthogonal to the
horizon and becomes null and Killing
over the horizon. At this point we can use any specific
 metric for
explicit computation and we illustrated the validity of
 Eq. (\ref{e}) by taking several examples.

% We wish to mention here that according to the proposition
%of the complex path method \cite{Kraus:1994}-\cite{Paddy3:2002},
%the path across the horizon is complex$-$there exists no real path
%or trajectory across the horizon. We followed this general 
%convention throughout this paper. However, there exist a different
%view of the path or the trajectory of the particle and 
%this may lead to
%some subtle issues \cite{Belinski}-\cite{Belinski:2009bc} which
%have not yet been fully resolved and needs further attention.

 The principle message of this work is the 
following. The semiclassical method provides us a
way through which we can treat the equations of motions
of different spin fields and compute the single particle 
emission probability or the Hawking temperature
for a stationary black hole
in an identical footing or manner.
\vskip 1cm

%%%%%%%%%%%%%%%%%%%%%%%%%%%%%%%%%%%%%%%%%%%%%%%%%%%%%%%%%%%
\section*{Acknowledgment}
I wish to sincerely acknowledge Amitabha Lahiri 
for useful discussions and encouragement. I also thank anonymous referees
for useful comments and questions. This work was supported by a fellowship from my Institution SNBNCBS.

%%%%%%%%%%%%%%%%%%%%%%%%%%%%%%%%%%%%%%%%%%%%%%%%%%%%%%%%%
\vskip 1cm
%%%%%%%%%%%%%%%%%%%%%%%%%%%%%%%%%%%%%%%%%%%%%%%%%%%%%%%%%%%%%%%
%                 BIBLIOGRAPHY
%%%%%%%%%%%%%%%%%%%%%%%%%%%%%%%%%%%%%%%%%%%%%%%%%%%%%%%%%%%%%%%%%%%%

\end{document}